\documentclass[]{agujournal2019}
\journalname{Journal of Advances in Modeling Earth Systems (JAMES)}
\usepackage{amsmath}

\usepackage{graphicx, float, multirow, tabu, hhline, overpic}
\usepackage{natbib}
\usepackage{amsfonts}
\usepackage{fancyhdr}
\usepackage{xcolor, soul}
\usepackage{booktabs}
\usepackage[noline]{algorithm2e}
\makeatletter
\renewcommand{\@algocf@capt@plain}{above}
\makeatother

\newcommand{\Lim}[1]{\lim\limits_{#1}\;}
\newcommand{\vv}[1]{{\mathbf #1}}
\renewcommand{\hline}{\noalign{\hrule height 1pt}}
\begin{document}
\title{Reservoir Computing as a Tool for Climate\\ Predictability
  Studies}
\authors{B.T. Nadiga\affil{1}}
\affiliation{1}{Los Alamos National Lab, Los Alamos, NM 87544}
\correspondingauthor{B. T. Nadiga}{balu@lanl.gov}

\begin{abstract}
  
Reduced-order dynamical models play a central role in developing our understanding of predictability of climate irrespective of whether we are dealing with the actual climate system or surrogate climate-models. In this context, the Linear-Inverse-Modeling (LIM) approach, by capturing a few essential interactions between dynamical components of the full system, has proven valuable in providing insights into predictability of the full system.

We demonstrate that Reservoir Computing (RC), a form of learning suitable for systems with chaotic dynamics, provides an alternative nonlinear approach that improves on the predictive skill of the LIM approach. We do this in the example setting of predicting sea-surface-temperature in the North Atlantic in the pre-industrial control simulation of a popular earth system model, the Community-Earth-System-Model so that we can compare the performance of the new RC based approach with the traditional LIM approach both when learning data is plentiful and when such data is more limited. The improved predictive skill of the RC approach over a wide range of conditions---larger number of retained EOF coefficients, extending well into the limited data regime, etc.---suggests that this machine-learning technique may have a use in climate predictability studies. While the possibility of developing a climate emulator---the ability to continue the evolution of the system on the attractor long after failing to be able to track the reference trajectory---is demonstrated in the Lorenz-63 system, it is suggested that further development of the RC approach may permit such uses of the new approach in more realistic predictability studies.

\end{abstract}

\section*{Plain Language Summary}
Because of the chaotic nature of the dynamics underlying many complex
systems such as weather and climate, evolution of ensembles of
trajectories have to be considered in order to produce future
predictions of such systems. An analysis of the dynamics of such
ensembles provides insights into mechanisms that make the system
predictable. Because comprehensive models of such complex systems are
very costly to run, reduced order dynamical models of them are a
useful tool in conducting such ensemble-based predictability
studies. We develop such a computationally-inexpensive reduced order
model using machine learning and show that it's predictive skill is
comparable to those of Linear Inverse Model (state of the art) when
training data is plentiful, but much better when such data is more
limited.  Consequently we think that this new method has  wide
applicability. Furthermore, given the nonlinear nature of the new
method, it has the potential to provide new insights into
predictability of complex systems.

\section{Introduction}
Following the pioneering work of Lorenz \citep[e.g.,][and
others]{lorenz1965study, lorenz1969three}, understanding to what
degree different aspects of the climate system are predictable has
become a foundational aspect of climate science. Indeed, since
predictability of climate can arise in two distinct ways, two kinds of
predictability are identified and are of interest
\citep{lorenz1975predictability}. In predictability of the first kind,
predictability arises from variability internal to the climate system
at various timescales when appropriately initialized and with
error-growth/decorrelation from observations being controlled by the
chaotic nature of the underlying dynamics. On the other hand, our
ability to anticipate the response of the climate system to
forcing external to the system, over a longer time scale gives rise to
predictability of the second kind. However, since our ability to
predict variations in climate as occurs due to internal/natural
variability is much less developed than our ability to predict the
response of the climate system to external forcing \cite[e.g.,
see][and references therein]{stocker2013climate}, we concern ourselves
with tools to address issues related to predictability of the first
kind in this article.

In the context of predictability of the first kind, because of the
chaotic nature of climate dynamics, it is necessary to consider the
evolution of an ensemble of trajectories be able to track the true
trajectory of the climate system. However, since we are limited to
observing the one climate system as it evolves, it is inevitable to
have to use models to make progress on understanding predictability of
the first kind. From the point of view of modeling, even if the
probabilistic evolution of the climate system can be formulated
correctly (e.g., as with the Liouville equation as in
\cite{ehrendorfer1994liouville}), the immense range of scales involved
and the computational complexity of such a formulation renders it
impractical and ensemble intergration of (severely) truncated models
is the only feasible alternative. In any such model, while on the one
hand, the difficulty of observationally estimating the state of the
climate system inevitably leads to initial condition errors, the
truncated representation of the climate system, on the other hand
leads to model error. Here, by model error we refer to   deficiencies
in the model such as insufficient resolution and   inaccurate
parameterizations of unresolved processes which lead to   the model's
inability to accurately simulate the delicate dynamical   balance of
processes that underlies both the mean state and the   modes of
variability of the climate system. Because of these errors---initial
condition errors and model errors---model predictions invariably
decorrelate rapidly from observations \citep[e.g.,
see][]{nadiga2019enhancing} and for this reason predictability studies
have had to largely focus on so-called ``perfect model'' scenarios
wherein {\em a} model trajectory is assumed to be the trajectory of
interest (``true'' trajectory) and the analysis of an ensemble of
model integrations forms the basis for characterizing predictability.

Although imperfect, extensive investments in climate modeling over the
past half a century have led to the development of a number of
comprehensive climate models and Earth System Models (ESMs), and, they
are proving invaluable in improving our understanding of various
details of the climate system including its variability. The
comprehensive nature of these models, however, renders them extremely
resource-intensive, computationally and otherwise. Even as
ensemble-based predictability studies with such models are beginning
to be performed at great expense, the model-specific nature of such
studies makes it difficult to relate the results of such studies to
the actual climate system. To wit, while studies of the Atlantic
Meridional Overturning Circulation (AMOC) in many models have found
decadal to multidecadal predictability \citep[e.g.,
see][]{griffies1997predictability, pohlmann2004estimating,
  collins2006interannual, hawkins2008potential, sevellec2008optimal},
these findings have tended to be disparate and model-specific, and
establishing the relevance of such findings to the real system has
been difficult.

A hierarchy of models, in terms of their complexity, has typically
been important to advancing our understanding of various phenomena in
climate science. For example, reduced-order dynamical models were
central to Lorenz's discovery of chaos and the ensuing body of work
related to predictability Lorenz \citep[e.g.,][and
others]{lorenz1965study, lorenz1969three}. As such, it should not be
surprising that reduced-order dynamical and empirical representations
of comprehensive climate models and observations continue to serve as
essential tools in our quest to further the understanding of
predictability of climate today. One such reduced-order dynamical
representation that has proved valuable is the Linear Inverse Model
(LIM). While more fundamentally rooted in the fluctuation-dissipation
theorem of equilibrium statistical mechanics, its typical usage in
climate science follows the work of Hasselmann
\citep{hasselmann1988pips} and Penland and co-workers \citep[e.g.,
see][]{penland1989random, penland1995optimal, farrell2001accurate}.
The LIM approach consists of a linear dynamical system
forced by white noise, where the linear dynamical operator and the
covariance of the noise in a reduced dimensional space---typically
Empirical Orthogonal Function (EOF) space are inferred from
data. Indeed, predictions using the LIM approach have been shown to be
skilful in a number of settings including the El Nino Southern
Oscillation (ENSO) \citep[e.g., see][]{penland1993prediction}, the
Pacific Decadal Oscillation (PDO) \citep[e.g.,
see][]{newman2007interannual, alexander2008forecasting}, and the
Atlantic Meridional Overturning Circulation (AMOC) \citep[e.g.,
see][]{hawkins2009decadal}.

Indeed, it is thought that the LIM approach is capable of capturing a
large fraction of the predictable signal in a variety of settings. To
wit, \cite{newman2017we}, in analyzing the predictability of tropical
sea surface temperature (SST) anomalies found that the forecast skill
of a LIM derived from
observed covariances of SST, sea surface height (SSH) and wind
fields was close to that of the (fully nonlinear and first-principles based) operational North
American Multi-Model Ensemble. The question then arises as to whether a linear approach,
as with a LIM, captures the entirety of the predictable signal or as to
whether there are predictable dynamics that are
nonlinear and therefore not capturable in a linear framework. A first
step towards exploring this issue consists of addressing the question
of whether the predictive skill of the well established LIM
methodology can be improved upon by further consideration of
non-linearity. This will be the focus of the present article.

The issue of improving on the skill of the LIM approach by
further consideration of nonlinearity itself has received substantial
attention, although largely in
the context of ENSO: Various authors (\cite[e.g.][and others; also see
\cite{kondrashov2018data}]{timmermann2001empirical,kondrashov2005hierarchy})
have considered nonlinear extensions of LIMs, but the benefits of such
extensions in terms of improved hindcast skill has been difficult to
establish and generalize.

Given the successes of data-driven methods in fields such as computer
vision, natural language processing, etc., over the past few decades,
an alternative approach to considering, nonlinearity has been to adapt
the artificial neural network (ANN) approach. While initial work in
the area \citep[e.g.][and
others]{grieger1994reconstruction,hsieh1998applying} was of an
exploratory nature, later studies starting with \citep{tang2000skill}
show improved skill of the (nonlinear) ANN approach in comparison to
linear methods \citep[e.g.,][and others]{ham2019deep}.

In the context of methods relevant to predictability studies where we
are interested in modeling temporal dynamics/processes, ANNs can be
categorized as either feedforward neural networks (FNN) or recurrent
neural networks (RNN). The use of feedforward networks in this setting
leverages the capability of such networks to approximate a continuous
function arbitrarily well \citep{cybenko1989approximation} to learn
the right hand side of an evolution operator. Here the evolution
operator is typically related to an underlying (unknown) partial differential
equation system if the system is considered in a physical domain or a
set of ordinary differential equations if the system is considered in a
modal domain and temporal evolution itself is achieved by the use of
traditional time integration schemes \citep[e.g.,
see][]{scher2019weather, weyn2019can,degennaro2018model,
  weyn2020improving}. Alternatively, an FNN can be used to directly
learn the future state at a fixed time increment given its recent
history. The latter RNN architecture, however, is
distinguished by the presence of cyclical connections in how the
neurons are connected \citep[e.g., see][]{lukovsevivcius2009reservoir}
and we concern ourselves with this architecture for the rest of this
article. Such an RNN while featuring deterministic dynamics that
transforms an input time series into an output time series through
nonlinear filters, is capable of exhibiting self-sustained temporal
dynamics.

In the context of RNNs, in response to various
shortcomings of mainstream RNN architectures, an important one of
which was the difficulty of training them, a new approach was proposed
independently by \cite{jaeger2001echo} and \cite{maass2002real}, and which
has subsequently come to be known as Reservoir Computing (RC). In RC,
an RNN that is randomly created and that remains unchanged during
training---the reservoir---is passively excited by the input signal
and maintains in the resevoir's state a nonlinear transformation of
the input history. Training then simply consists of using linear
regression to obtain the weights that best give the desired output
signal as a linear combination of the input and the resevoir state. In
particular, since RC has outperformed other methods of nonlinear
system identification, prediction, and classification in the context
of chaotic dynamics \citep[e.g.,][and others]{jaeger2004harnessing,
  pathak2018model}, we consider its use as a tool in climate predictability
studies, while noting recent use of RC methods in the context of
weather prediction \citep{arcomano2020machine}.

The rest of the article is structured as follows: In the next section
we present the details of the problem we consider. Following that, in
section 3, we present the Linear Inverse Model approach to the problem
and discuss various details of the approach. After presenting details
of the reservoir computing approach to the problem in section 4, we
compare the results of this approach with the results of the LIM
approach in section 5. A discussion of the results and some additional
experiments using the Lorenz-63 system towards gaining some insight
into the RC method and its performance in the climate setting are presented in
section 6. A few implications of the study and a short discussion of
the pros and cons of the new methodology then concludes the article.

\section{Predictability of the North Atlantic Sea Surface Temperature}
The utility of skilful near-term (subseasonal to decadal) predictions
of regional climate is manifold and range from assessing societal and
ecological impacts of a changing climate \cite[e.g.,
see][]{national2008ecological,adger2010social} to planning and
managing infrastructure \cite[e.g., see][]{wilbanks2014climate} to
insurance and risk management \cite[e.g., see][]{mills2005insurance}
to adapt to such changes.
In the framework of comprehensive climate
and earth system models, a host of reasons, including the scientific
challenges involved in being able to estimate the state of the climate
system with sufficient accuracy and the complex, multiscale and
chaotic dynamical nature of the climate system which complicates the
process of accounting for uncertainty in the future evolution of
errors in the initial state estimate, make predictions of the first
kind more difficult than being able to model the response of the
climate system to secular changes in external forcing such as due to
greenhouse gases (e.g., see \cite{meehl2009decadal,
  meehl2014decadal}). As such, our ability to produce longer term
projections, projections that are contolled by external forcing
related predictability, is better developed than our ability to
produce near-term (subseasonal to decadal) predictions---predictions that are
(increasingly) controlled by natural variability related predictability (as
the prediction lead time decreases).  It should,
however, also be noted that the response of the climate system to
external forcing can be/is modulated by natural variability, leading
to the response to external forcing being amplified or mitigated on
certain time scales of natural variability.

Remaining in the framework of comprehensive climate models, while
initialized predictions of climate seek to augment the external
forcing related predictability that is realized in uninitialized long
term projections by predictability related to natural variability,
there are a number of issues that remain to be resolved before such
initialized predictions are skillful \citep[e.g.,
see][]{kim2012evaluation, kharin2012statistical, sanchez2016drift,
  nadiga2019enhancing}. For example, in many such models observation
based initialization in the presence of model bias leads to a rather
rapid departure of the initialized prediction trajectory from
observations necessitating post-processing of the predictions before
they can show any skill at all. For these reasons, we concern
ourselves with a statistical approach to the problem of near-term
prediction of climate presently.

We consider the variability of SST in the North Atlantic over the last
800 years of the pre-industrial control (piControl; a simulation in
which external forcing is held fixed) simulation of the Community
Earth System Model \citep[CESM2;][]{danabasoglu2020community} as part
of the sixth phase of the Coupled Model Intercomparison Project (CMIP6). CESM2
is a global coupled ocean-atmosphere-land-land ice model and the
piControl simulation we consider uses the Community Atmosphere Model
(CAM6) and the Parallel Ocean Program (POP2), and at a nominal 1$^o$
horizontal resolution in both the atmosphere and the ocean; the reader
is referred to \cite{danabasoglu2020community} for details.  This data
is publicly available from the CMIP archive at
https://esgf-node.llnl.gov/projects/cmip6 and its mirrors.

\begin{figure}
  \includegraphics[width=\textwidth]{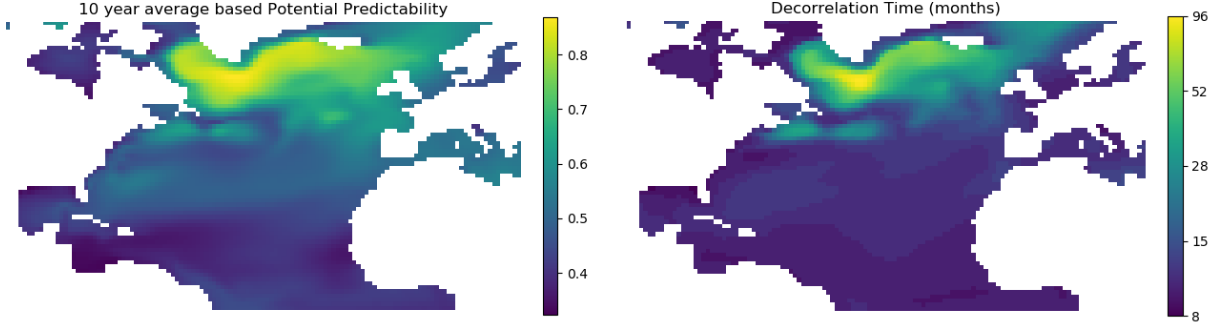}
  \caption{Two measures of predictability of SST in the North
    Atlantic. On the left is shown potential predictability computed
    as the pointwise ratio of the standard deviation of the ten year
    average to the standard deviation of the one year average. On the
    right is shown a measure of the decorrelation time computed as the
    pointwise e-folding time of the auto-correlation function. The
    time is shown in months on the colorbar.}
  \label{decor}
\end{figure}

Two measures of predictability of SST in the North Atlantic are shown
in Fig.~\ref{decor}.  On the left is potential predictability and on
the right is a decorrelation time. Potential predictability is defined
as the ratio of the standard deviation of the N-year average of SST
(in the current context) to the corresponding standard deviation of
the 1-year average \citep{boer2004long}. As such, this measure assumes
that variability on the longer (N$>$1) timescales is the signal related
to potentially predictable processes (such as due to slow
deterministic ocean dynamics) while that on the (faster) interannual
scale is noise (attributable to processes such as chaotic internal
variability). Potential predictability using a ten year average is
shown in the left panel of Fig.~\ref{decor}.

Various definitions of decorrelation time ($\tau_D$) are possible
\citep[e.g., see][and references therein]{von2001statistical} and
three were considered:
\begin{eqnarray}
\tau^{(1)}_D =& 1 + 2 \sum_{k=1}^{\infty} \rho(k),\cr
\tau^{(2)}_D =& 1 + 2 \sum_{k=1}^{\infty} \rho^2(k),
\end{eqnarray}  
and simply the e-folding time of the auto-correlation function
$\rho(k)$ ($\tau^{(3)}_D = k$ at which $\rho(k)$ falls below $1/e$ for
the first time). The computed values of $\tau^{(2)}_D$ and
$\tau^{(3)}_D$ showed strong similarity with the former greater than
the latter by about 20\%, whereas $\tau^{(1)}_D$ showed a much larger
range of values. For brevity and as a more conservative estimate of
predictability, $\tau^{(3)}_D$ is shown in the right panel of
Fig.~\ref{decor}. In this figure, the color bar for the decorrelation
time is shown in months .

Both measures of predictability in Fig.~\ref{decor} indicate that the
more northern regions of the North Atlantic are more predictable, and that
in these regions, predictability may extend to times as long as about eight years. We
note that this finding is consistent with previous analysis of the variability of SST of the
global ocean, both in observations and climate models that has identified
the North Atlantic as a region that possesses significant
predictability \cite[][and others]{delworth2000observed, boer2004long,
hawkins2011evaluating}. As such, we consider the spatio-temporal
variability of SST in the North
Atlantic for our example setting.

\begin{figure}
  \includegraphics[width=\textwidth]{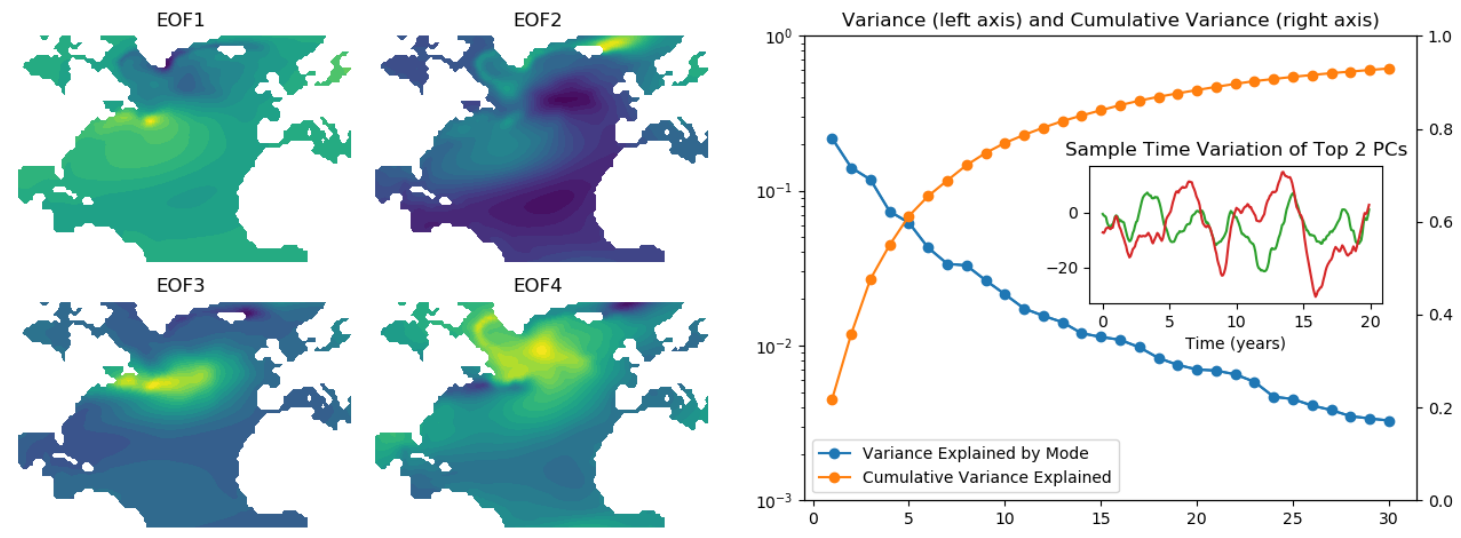}
  \caption{The leading EOFs of North Atlantic SST are shown on the
    left. Scale information is omitted since the EOFs are
    normalized. The main plot in the right panel shows the fraction of
    variance captured by individual modes (left axis) and the
    cumulative variance captured as a function of the mode number
    (right axis). The inset in the right panel shows
    time variations of the two leading principal components over a
    period of 20 years.}
  \label{eofs}
\end{figure}


\section{Linear Inverse Modeling}
Since we are interested in predicting interannual variations of the SST
based on data from the CESM2 simulation described above, we
consider a twelve month moving-window average of the monthly SST
field in what follows. The measures of predictability considered in
the previous section were estimated in a spatially-local, pointwise
fashion. A commonly used approach to further consider the effects of
spatial covariance is that of empirial orthogonal
function (EOF) analysis (equivalently principal component analysis
PCA) \citep[e.g., see][and references therein]{von2001statistical}.
Furthermore, a truncated EOF basis also serves as an  effective
strategy to reduce the dimension of the system under consideration:
\begin{eqnarray}
  T(\vv x, t) = \sum_{k=1}^N  e_k(\vv x)  y_k(t),
\end{eqnarray}
where T is SST, $e_k(\vv x)$ is the $k-th$ EOF, $y_k(t)$ its
corresponding time coefficient or principal component, and where $N$
the number of EOFs retained is much smaller than the number of spatial
locations $\vv x$.

The spatial pattern of the first four EOFs on performing the analysis
using the full record of 800 years is shown in the left panel of
Fig.~\ref{eofs}; scaling information/colorbars are omitted since the
EOF patterns are normalized. The main plot in the right panel of
Fig.~\ref{eofs} shows the fraction of variance captured by individual
modes and the cumulative variance captured as a function of the mode
number (using two separate vertical axes on the left and right
respectively). While the first four modes capture about 50\% of the
variance, the first thirty modes capture about 95\% of the variance.
To give the reader a feel for the nature of temporal variability in
the EOF basis, the inset in the right panel of Fig.~\ref{eofs} shows
time variations of the two leading principal components (time
coefficients) over a period of 20 years.

In its typical usage in climate science, the LIM approach consists of
considering the evolution of the leading principal components as a
linear dynamical system that is forced by white noise \citep[e.g.,
see][]{hasselmann1988pips, penland1989random, penland1995optimal, farrell2001accurate}:
\begin{eqnarray}
  \frac{d\vv y}{dt} = \vv B\vv y + \boldsymbol\xi,
  \label{limeq}
\end{eqnarray}  
where $\vv y$ is the state vector, i.e., the truncated vector of
principle components. In (\ref{limeq}) $\vv B$ is a constant
deterministic matrix determined as:
\begin{eqnarray}
  \vv B = \frac{1}{\tau_0} \ln\left(\vv C_{\tau_0} \vv C_0^{-1}\right),
  \label{limeqB}
\end{eqnarray}
where $\vv C_{\tau_0} = \left<\vv y(t+\tau_0) \vv y^T(t)\right>$ is
the lagged covariance at a chosen lag $\tau_0$, $\vv C_0 = \left<\vv
  y(t) \vv y^T(t)\right>$ is the (zero-lag) covariance, 
and $\boldsymbol\xi$ is a (vector) white noise process with
a covariance matrix $\vv Q$ ($=\left<\boldsymbol\xi\boldsymbol\xi^T\right>$) that is
determined from:
\begin{eqnarray}
  \vv B\vv C_0 + \vv C_0 \vv B^T + \vv Q = 0.
\end{eqnarray}
See \cite{penland1989random} for details. The LIM (\ref{limeq}), can
then be used to predict $\vv y$ at lead time $\tau$ as
\begin{eqnarray}
  \vv y_L(t + \tau) = \exp(\tau \vv B)\vv y(t) + \exp(\tau \vv
  B)\int_t^{t+\tau}\exp(-t' \vv B)\boldsymbol\xi(t') dt'.
\end{eqnarray}
However, if we are only interested in the ensemble average over
realizations of the noise process, then 
\begin{eqnarray}
  \bar{\vv y}_L(t + \tau) = \exp(\tau \vv B)\vv y(t).
  \label{avgLIM}
\end{eqnarray}
In the rest of the article, we'll only consider the ensemble average
prediction and drop the overbar for convenience:
\begin{eqnarray}
  \vv y_L(t + \tau) = \exp(\tau \vv B)\vv y(t) = \vv P_\tau \vv y(t).
  \label{limpred}
\end{eqnarray}
Here it is important to note that $\vv P_\tau$, the propagator of the
state vector $\vv y$ over a period $\tau$ has to tend to zero at long
lead times:
\begin{eqnarray}
\Lim{\tau\rightarrow \infty}\vv P_\tau = 0
\end{eqnarray}
or equivalently that the real part of all of the eigenvalues of
$\vv B$ have to be negative for the LIM to be useful.

Finally, if, the true dynamics of the system were indeed linear, the
expected mean square error of the prediction $\vv y_L(t + \tau)$ at
lead time $\tau$ is (related to the covariance of the (vector) noise
process and) given by
\begin{eqnarray}
  \epsilon_{ifL}(\tau) = \frac{\hbox{tr}\left(\vv C_0 - \vv P_\tau \vv C_0
  \vv P^T_\tau\right)}{\hbox{tr}(\vv C_0)},
  \label{iflerr}
\end{eqnarray}
where tr($\cdot$) is the trace of the relevant matrix. However, since the
actual evolution of the principal components of SST is likely
nonlinear, a comparison of the actual error of the ensemble-averaged prediction 
$\vv y_L(t + \tau)$ in (\ref{limpred}) to the theoretical estimate of
the error if the actual evolution were to be linear in (\ref{iflerr})
can be used as a measure of nonlinearity in the actual evolution.

We end this section by noting a couple of alternative/complementary
views of the LIM approach. While we presented the LIM approach
here as a somewhat ad-hoc ``reduced order modeling''
approach, the methodology has deeper roots in statistical mechanics
and dynamical systems theory. Indeed the SST prediction problem
can be recast as one seeking, in effect, a global space-time response
of a statistically steady surface ocean from a dynamical
systems perspective. In this context, the Fluctuation Dissipation
Theorem (FDT) of statistical mechanics relates the linear response of
the system to certain space-time correlation functions of the
undisturbed system \cite[e.g., see][]{leith1975climate,
  gritsun2007climate, abramov2007blended}: The surface ocean is
continuously experiencing small forcings of numerous different types,
and these generate corresponding fluctuation responses. Therefore,
given a sufficiently detailed space-time map of these correlations,
the FDT allows the response to a given force to be extracted and
resolved, leading exactly to the LIM formulation described above.

Alternatively, when used as a data-driven approach to approximate the
evolution of an underlying nonlinear dynamical system, the LIM
approach, in common with the Dynamical Mode Decomposition (DMD)
approach \cite[e.g., see][]{tu2014dynamic}, arises in the context of
spectral analysis of the Koopman operator---a linear but
infinite-dimensional operator whose modes and eigenvalues capture the
evolution of observables describing any dynamical system, even
nonlinear ones \cite[e.g., see][]{tu2014dynamic}. In this context, (a)
the propagator $P_\tau$ in (\ref{limpred}) is the Koopman operator, on
using the state vector itself as the observable. However, since the
Koopman operator acts on  functions rather than on the state itself,
the Koopman operator steps the observation operator forward in
time, and since the observation
operator here is the identity operator, in effect the state is stepped
forward in time, and (b)  rather than thinking of $P_\tau$
as arising from a linearization of the underlying dynamics, it is better
thought of as an average of the nonlinear dynamics evaluated over an
ensemble of snapshots \citep[e.g.,
see][]{blumenthal1991predictability}.
\begin{figure}
  \includegraphics[width=\textwidth]{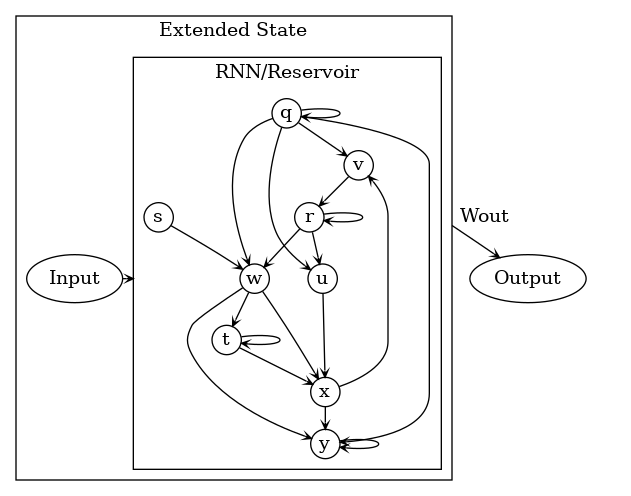}
  \caption{A schematic of the reservoir computing architecture. The
    connectivity between the neurons in the reservoir is random and
    training is limited to determining $\vv W_{out}$ using (a
    regularized form of) linear
    regression over the training data. In the prediction phase, the
    output becomes the input for the next time step.}
  \label{singleres}
\end{figure}

\section{Reservoir Computing}
As mentioned in the introduction, Reservoir Computing is a form of
Recurrent Neural Network (RNN) in which a randomly created reservoir
stores a set of nonlinear transformations of the input signal and in
which training consists of using linear regression to obtain the
weights that best give the desired output signal as a linear
combination of the input and the resevoir state \citep[e.g., see][and
references therein]{lukovsevivcius2009reservoir}. 

Following the development in \cite{lukovsevivcius2009reservoir}, in
general, an RNN may be thought of as mapping a (vector) input time
series $\vv u(n)$ to a (vector) output time series $\vv y(n)$ with
$n=1,\ldots,T$, and e.g., as shown in schematic
Fig.~\ref{singleres}. We note here that when the dimensionality of the
problem is reduced, as we do by using EOFs here, $\vv y(n)$ comprises
of the time evolving coefficients of the retained modes.  However, if
such dimension reduction techniques were not employed, while
$\vv y(n)$ would comprise of the temporally evolving values of
relevant variables at different spatial locations, spatial-locality of
the governing dynamics would lead us to consider separate reservoirs
for each of the domain-decomposed neighborhoods with further exchange
of (boundary or halo) information between adjoining reservoirs at regular time
intervals.  For the kinds of applications we consider in this article,
it is natural to consider $\vv u(n) = \vv y(n-1)$ (autonomous
dynamics). (But for the difference that augmenting the input vector $\vv u(n)$ by a
unit entry is a convenient way of including bias in both the reservoir
update (see (\ref{resevolve})) and in the readout (see (\ref{readout})).
As such, input and output in this article refer to the evolving state
vector $\vv y$ at two subsequent time steps and where the time step is
fixed at a month. Note that we refer to the vector $\vv y$ as the
state vector, the vector $\vv r$ as the resevoir state (vector), and
the combined vector \{$\vv r ,  \vv y$\} as the extended state
(vector). Finally, when we mention {\em input} in the rest of the
article, we refer to $\vv u(n)$ in the more general (non-autonomous)
setting or to $\vv y(n)$ in the present (autonomous) setting.

\subsection{Nonlinear Kernels and Linear Readout}
Indeed, to investigate the possibility of
improving on the {\em Linear} inverse modeling approach of the
previous section, we are interested in going beyond $\vv y(n+1) = \vv W
\vv y(n)$ where (and in the rest of the article) $\vv W$ is a constant
weight matrix. A generic approach to nonlinear modeling is to first consider
transformation of the input $\vv y(n)$ into a high-dimensional set of
nonlinear features $\vv r(n)$ and then apply linear techniques such as
linear regression to obtain a nonlinear model. Following this approach
in the present context leads to the nonlinear model
\begin{eqnarray}
  \vv y(n+1) = \vv W_{out} \tilde{\vv r}(n).
  \label{readout}
\end{eqnarray}  
Here, again, $\vv W_{out}$ is a constant weight matrix of dimension
$N_y\times N_{\tilde r}$, but with $N_{\tilde r}\gg N_y$ typically,
and where $\tilde{\vv r}$ is the combined vector of $\vv r(n)$, the set
of nonlinear features and $\vv y(n)$, the input: $\tilde{\vv r} = \left\{\vv r(n)
  ,  \vv y(n)\right\}$, with $N_{\tilde r} = N_r +
N_u$. 

\subsection{Recurrent Neural Networks}\label{RNN}
The storage of the nonlinear expansion/transformations of the input
with memory in an RNN is given by the evolution of the RNN state as
\begin{eqnarray}
  \vv r(n) = f\left(
  \vv W\vv r(n-1) + 
  \vv W_{in} \vv u(n)
  \right), n=1,\ldots,T,
  \label{resevolve1}
\end{eqnarray}
where $\vv r(n)$ and $\vv u(n)$ are respectively the vector of RNN
states and the vector of inputs at time step $n$, and $f(\cdot)$ is
the activation function applied element-wise.  In (\ref{resevolve1}),
$\vv W_{in}$ and $\vv W$ are both weight matrices with dimensions
$N_r\times N_u$ and $N_r\times N_r$ respectively. We note here that
the bias term in (\ref{resevolve1}) is included by adding a unit
entry to the input vector.

Indeed, a yet-simpler approximation that foregoes the connectivity
between the neurons in the reservoir in (\ref{resevolve1}) and relies
entirely on random features to construct kernel machines has been
considered by various authors \citep[e.g., see][]{rahimi2008random,
  gottwald2020supervised}.

Finally, we note that, to account for certain details of the system
such as the data frequency, (e.g., on expressing data frequency in
terms of a characteristic decorrelation time of the system), it is
useful to generalize (\ref{resevolve1}) by introducing a so-called
``leakage'' parameter $\alpha_{lk}$:
\begin{eqnarray}
  \vv r(n) = \left(1-\alpha_{lk}\right)\vv r(n-1) + \alpha_{lk} f\left(
  \vv W\vv r(n-1) + 
  \vv W_{in} \vv u(n)
  \right), n=1,\ldots,T.
  \label{resevolve}
\end{eqnarray}

\subsection{Constancy of RNN Weights in Reservoir Computing}
While in most forms of RNNs, training is accomplished by iteratively
adapting all weights involved (e.g., $\vv W_{in}$, $\vv W$, and
$\vv W_{out}$), in reservoir computing, all weight vectors except
$\vv W_{out}$, the weights associated with the output layer are set
randomly and held constant, and training is accomplished by setting
$\vv W_{out}$ using linear regression. In so doing, RC methods differ
from other RNNs in partitioning off the recurrent (bulk) parts of the
network as a dynamic reservoir of nonlinear transformations of the
input history and isolating the training to readout parts of the
network that not only do not contain any recurrences but are also
usually linear.

Notwithstanding the vastly simplified nature of RC in comparison to
other RNN methodologies, it still remains that setting up a ``good''
reservoir for a particular problem or for a class of problems is still
poorly understood. In response to this, and again in common with other
ML techniques, there is a proliferation of RC methodologies. As such,
and again in common with the use of most machine learning techniques,
a certain degree of trial and error using accumulated
experience/heuristics in the field as a guide is essential to
successfully using the RC approach.

\subsection{Echo State Networks: Weighted Sum and Nonlinearity}
Given our primary interest in investigating the effects of considering
nonlinearity in prediction problems related to climate and comparing
prediction skill against the well established LIM approach, we focus
on simple but robust formulations of the RC method. As such, we focus
on Echo State Networks (ESN) \citep[e.g., see][and references therein]{jaeger2007special},
networks that essentially embody the ``weighted sum and nonlinearity''
aspect of RCs seen in (\ref{readout}) and (\ref{resevolve}) and
training is reduced to linear regression.

\subsection{Echo State Property and Scaling of Input Connections}
Analogous to the LIM requirement that $\vv P_\tau\rightarrow 0$ as
$\tau\rightarrow \infty$ for it to be useful, a requirement for ESNs
to work is that they should posses the echo state property
\citep{jaeger2001echo} whereby the effect of a previous reservoir
state $\vv r(n)$ and the corresponding input $\vv u(n)$ on a future
state $\vv r(n+k)$ should vanish as $k\rightarrow\infty$. While this
is usually assured when the spectral radius of $\vv W$ is less than
unity, this requirement on the spectral radius is quite often not
necessary \citep[e.g., see][and references
therein]{jaeger2007special}. Here, the spectral radius of $\vv W$ is
given by the eigenvalue of $\vv W$ that has the largest magnitude.
Specifically, a random matrix of dimension $N_r\times N_r$ with
sparsity $S$, $\widehat{\vv W}$, was generated using a unit normal distribution centered
at zero ($N(0,1)$). Thereafter, the individual entries of the matrix were
rescaled by the magnitude of the eigenvalue with the largest magnitude
($\lambda_{max}$) to obtain the
random reservoir connectivity matrix $\vv W$ with specified spectral 
radius ($\rho$) as
\begin{equation}
  \vv W = \frac{\rho}{\lambda_{max}}\widehat{\vv W}\left(N(0,1)\right).
  \label{connectivity}
\end{equation}

While the spectral radius of the reservoir connectivity matrix $\vv W$
is related to the {\em echo state property} of the reservoir, how
strongly the reservoir dynamics is driven by current value of the
state vector ($\vv u(n)\equiv \vv y(n-1)$) itself is related to
magnitude of the entries of $\vv W_{in}$: Again, a random matrix of
dimension $N_r\times N_u$ is generated using a unit normal
distribution and then multiplied elementwise by a factor $\alpha^2_{in}$ to
obtain $\vv W_{in}$:
\begin{equation}
  \vv W_{in} = \alpha^2_{in}\widehat{\vv W}_{in}\left(N(0,1)\right)
  \label{alphain}
\end{equation}  

\subsection{Training of Linear Readout using Ridge Regression}
Having set the internal connectivity of the reservoir and its
connectivity to the input, it remains to determine 
$\vv W_{out}$ before the network can be tested and used to predict the
future evolution of the system. This process of {\em training} the
readout (\ref{readout}) is achieved as follows: Given the (standarized) state vector
$y$ over a training period $\vv y(n), n=1,\ldots,T_{tr}+1$, the
reservoir state $r(n), n=1,\ldots,T_{tr}$ is computed using
(\ref{resevolve}) after initializing the reservoir state
($\vv r(0) = 0$). After an initial washout period $T_{wo}$, the
extended state of the system (i.e., the state of the resevoir and the
input state vector) is collected together in a collection matrix
$\vv R$ of dimension
$ N_{\tilde r} \times \left(T_{tr}-T_{wo}\right)$. Over the same
period, a matrix $\vv Y$ of dimension
$ N_y\times\left(T_{tr}-T_{wo}\right)$ that comprises the
one-step-advanced state vector $y(n+1), n=T_{wo}+1, \ldots, T_{tr}$ is
formed. Then, writing the readout equation (\ref{readout}) in a matrix
form leads to
\begin{equation}
  \vv  Y = \vv W_{out}\vv R
  \label{train}
\end{equation}  
which constitutes a linear regression problem to determine a $\vv
W_{out}$ (that typically minimizes the quadratic error between the two
sides of (\ref{train})). While the Moore-Penrose pseudoinverse of $\vv R$ provides a
direct and numerically-stable solution to the linear regression
problem, it can be expensive memory-wise for long training periods
\cite[e.g., see.][]{lukovsevivcius2009reservoir}. Ridge regression
(equivalently the Tikhonov regularization) provides a solution to
(\ref{train}) 
\begin{equation}
\vv W_{out} =   \vv  Y \vv R^T\left(\vv R\vv R^T + \alpha^2\vv I\right)^{-1}
  \label{ridge}
\end{equation}  
that is also numerically stable, but is also usually faster
than the pseudoinverse. Furthermore regularization, through the
$\alpha^2$ term, in (\ref{ridge}), by 
limiting the magnitude of entries in $\vv W_{out}$ serves to mitigate
sensitivity to noise and overfitting
\citep{lukovsevivcius2009reservoir}, and making ridge regression the
solution of choice.

\subsection{Making Predictions with Reservoir Computing}
Predictions can be made with the trained reservoir by continuing the
update of the system beyond the end of training ($n=T_{tr}$). In
particular, the reservoir is updated according to (\ref{resevolve}),
the one-step prediction is obtained from (\ref{readout}), the two-step
prediction is obtained by feeding the one-step prediction as input and
so on. Considering that the weights $\vv W$ and $\vv W_{in}$ are
random, we average over $N_{rand}$ realizations of the network to
obtain the averaged prediction, much in the sense of considering the
average prediction of a LIM over multiple realizations of the noise
process, as in (\ref{avgLIM}). Again, for convenience and as with the
LIM prediction, we generally omit the overbar for the prediction
$\vv y_R(T_{tr}+n), n=1,\ldots$

\subsection{The Reservoir Computing Algorithm and Parameter Values}

\begin{algorithm}
   \Indm
  \KwData{Evolution of the state vector
    $\vv y$ over the training period $\vv y(n), n=1,\ldots,T_{tr}$}
  \KwResult{Prediction $\vv y_R(T_{tr}+
    \left(k-1\right)T_{skip}+n),\; ;k=1,\ldots, N_{st},\; n=1,\ldots,T_{pred}$}
  \For{$\xi$ = 1 \KwTo $N_{rand}$}
  {
    \vspace*{.2cm}
  {\bf Build Reservoir:} Using values from Table 1 and description in
  this section\\
  \Indp
  Set up reservoir connectivity matrix $\vv W$ using (\ref{connectivity})\\
  Set up input to reservoir connection matrix $\vv W_{in}$ using (\ref{alphain})\\
  \Indm
  {\bf Train Readout:}\\
  \Indp
  Initialize reservoir state $\vv r(0) = 0$\\
  Form input ($\vv u$) and output ($\vv y$) vectors over training period\\
  Step reservoir over training period using (\ref{resevolve})\\
  Collect extended state (reservoir state and state vector) \{$\vv r , 
  \vv y$\} over the training period, after discarding a short burn-in/wash-out
  period $T_{wo}$, in the form a matrix $\vv R$,
  and target one-step-prediction in the form a matrix $\vv Y$\\
  Obtain $\vv W_{out}$ using ridge regression (\ref{ridge})\\
   \Indm
   {\bf Predict future evolution:}\\
   \Indp
   \For{$k=1$ \KwTo $N_{st}$}{    \vspace*{.2cm}
   $T_{st} = T_{tr} + \left(k-1\right)T_{skip}$\\
   (Re)compute reservoir evolution to $T_{st}$ using data\\
  \For{$n=1$ \KwTo $T_{pred}$}{    \vspace*{.2cm}
    Update reservoir state $\vv r(T_{st}+n)$ using (\ref{resevolve})\\
    Obtain prediction $\vv y(T_{st}+n+1)$ using (\ref{readout})\\
    Use prediction as input for next time step: $\vv u(T_{st}+n+1)
   \leftarrow \vv y(T_{st}+n+1)$\\
 }
 \Indm
}
}
$\vv y_R \leftarrow \left<\vv y(T_{tr} +
  \left(k-1\right)T_{skip}+n)\right>_\xi, \;k=1,\ldots, N_{st},\;
n=1,\ldots, T_{pred}$\\\vspace*{0.4cm}
\caption{A recap of the reservoir computing methodology used in the article.}
  \end{algorithm}

\begin{table}
\centering
\caption{Parameter values and other details of the reservoir computing
method used.}
\label{table}
\begin{tabular}{ |r|l| }
  \hline
  \multicolumn{2}{|c|}{Details of Reservoir Architecture }\\
  \hline
  Type& Echo State Network\\
  \hline
Topology &Random \\
\hline
Number of neurons in reservoir, $N_r$ &200 \\
\hline
Spectral radius ($\rho$ in (\ref{connectivity})) &1 \\
\hline
Sparsity of reservoir connectivity matrix&0 \\
\hline
Scale for input connections ($\alpha^2_{in}$ in (\ref{alphain})) &
                                                                   $10^{-2}$\\
  \hline
  Initial washout period $T_{wo}$ in months & min(100, 0.1*$T_{tr}$)\\
  \hline
  Ridge regression coefficient ($\alpha^2$ in (\ref{ridge})) &
                                                               0.1\\
  \hline
Activation function $f$ &Symmetric tanh \\
\hline
  Distribution for random weight matrices &Centered unit normal \\
  \hline
  Number of instances to average over, $N_{rand}$&32\\
\hline
\end{tabular}
\end{table}

We note that in using RC in this study, our emphasis is on simplicity
rather than the best performance. As such we forego both, the many
other variations in architecture that are possible with RC, and extensive tuning of
the reservoir for performance. A reasonable set of parameters that was
found in an initial exploration of the methodology in one of the
cases constitutes for the most part the single set of parameters that are
used for all the cases. A recap of the algorithm in
Algo.~1 and a listing of the parameter values used in Table.~1
completes the description of the reservoir computing approach we adopt
and we next proceed to compare the performance of this simple
(nonlinear) network against the linear inverse modeling approach.

\section{Comparison of Skill of Reservoir Computing and LIM
  Approaches}
The requirement that the real parts of all of the eigenvalues of $\vv
B$ in (\ref{limeq}) be negative for the LIM approach to be useful
can be particularly problematic when the data record is short. When
this has been the case, a variety of ad-hoc fixes have been adopted by
practitioners to ensure the stability
requirement. For example, \cite{hawkins2011evaluating} satisfy the
requirement by limiting the number of EOFs considered (to 7 in their
context). Other ways of ensuring the stability of LIM is by
increasing the level of smoothing of the data, etc.

\subsection{Learning from Long Runs of Data}
In the first setting that we will consider, we want to avoid the need
for such ad-hoc fixes or restrictions when using the LIM
approach. As such, we use a {\em long} training period: That is we use
the initial 60\% of the 800 year record as training data.

The determination of the (constant) matrix $\vv B$ of the LIM
(\ref{limeq}) using (\ref{limeqB}) involves the choice of a lag
$\tau_0$, and the dependence of $\vv B$ on $\tau_0$ (also suggestive
of nonlinearity in the behavior of the system), has to be addressed.
We examined the dependence of error (NRMSE) and anomaly correlation
coefficient (ACC) on the choice of lag $\tau_0$ and picked the value
of $\tau_0$ that produced minimum error and maximum correlation. On
examining error and correlation for
$\tau_0\in \{1, 2, 3, 4, 6, 12, 18, 24\}$ months, we found the
performance to be best at a lag of six months, while noting that the
performance with a lag of twelve months was only slightly worse. As
such, a lag $\tau_0$ of six months was chosen and we call this
'bestLIM' for brevity. The choice of lag is determined by evaluating
the skill of predictions over a 20\%
validation segment of the data. After the lag is thus fixed, training
is performed a second time over the 80\% of data (comprising the
training and validation segments) and testing is performed over the
remaining 20\%.

Since the parameters in the RC approach are fixed at the values
specified in the table, in this setting, the RC system is trained over
the same  80\% of data (comprising the
training and validation segments) and testing is performed over the
remaining 20\%. That is a $\vv
W_{out}$ is found through linear regression as described in the
previous section and held fixed, while predictions starting at a
number of different start dates, $T_{skip}$ months apart, are initialized using the correct
state of the reservoir for the start date.
\newcommand{\prfx}{figs-submit2/200_480yr_18_1919_r32la:1_le:1_sp:1_ri:0.1_sp:0.5_wi:0.01}
\begin{figure}
  \includegraphics[width=\textwidth]{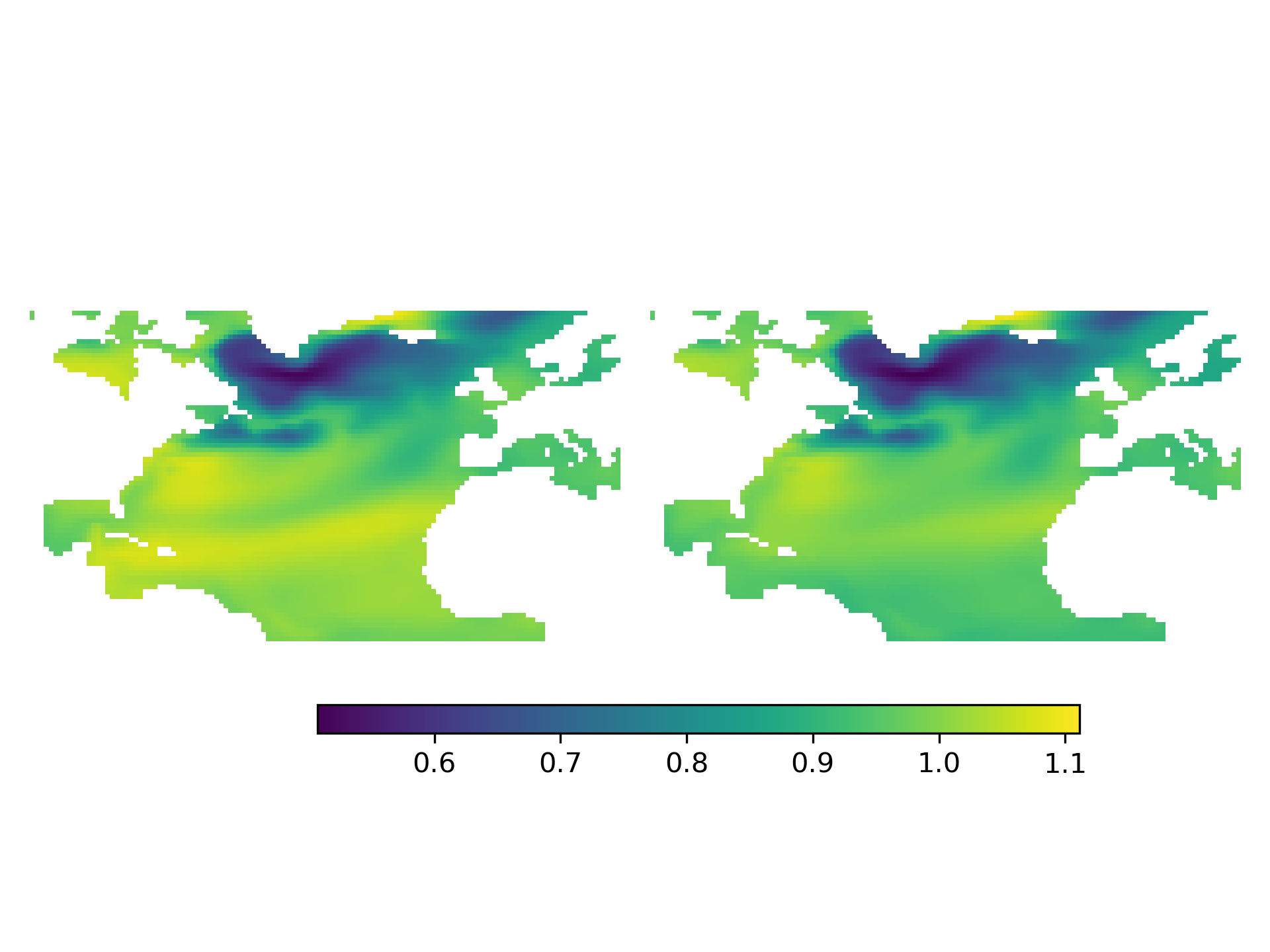}
  \caption{Spatial distribution of the normalized root mean square
    error at a prediction lead time of one year, averaged over
    ($N_{st}$=213) prediction starts. The LIM result is shown on the
    left and the RC result on the right. While the overall patterns of
    error are similar (lower errors in the sub-polar gyre, etc.), the
    RC methodology is seen to display consistently smaller errors
    (e.g., note the brighter yellow in the LIM result midway between
    Florida and Africa and further South).}
  \label{xyerrlong}
\end{figure}
\begin{figure}
  \includegraphics[width=\textwidth]{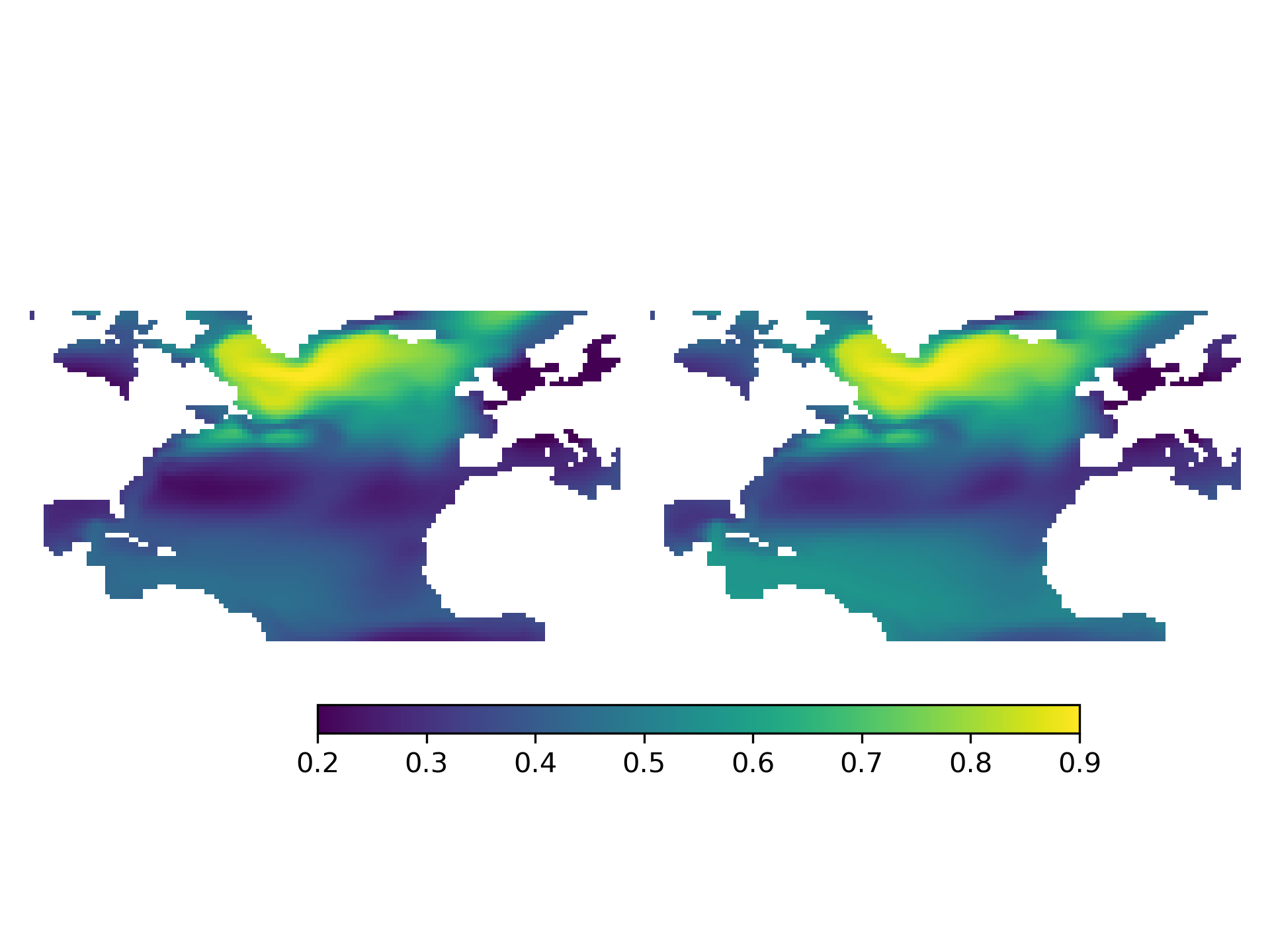}
  \caption{Anomaly correlation coefficient between the reference CESM2
  SST and the LIM predictions (left) and the RC
  predictions (right) at the same prediction lead time of one year,
  computed over ($N_{st}$=213) prediction starts. Similar to the
  comparisons of errors in Fig.~\ref{xyerrlong}, while the overall
  patterns are again the same in the two approaches, the RC
  methodology displays consistently higher correlation skill (e.g.,
  again focussing on the region midway between Florida and Africa and
  further South).}
  \label{xyacclong}
\end{figure}

\begin{figure}
  \includegraphics[width=\textwidth]{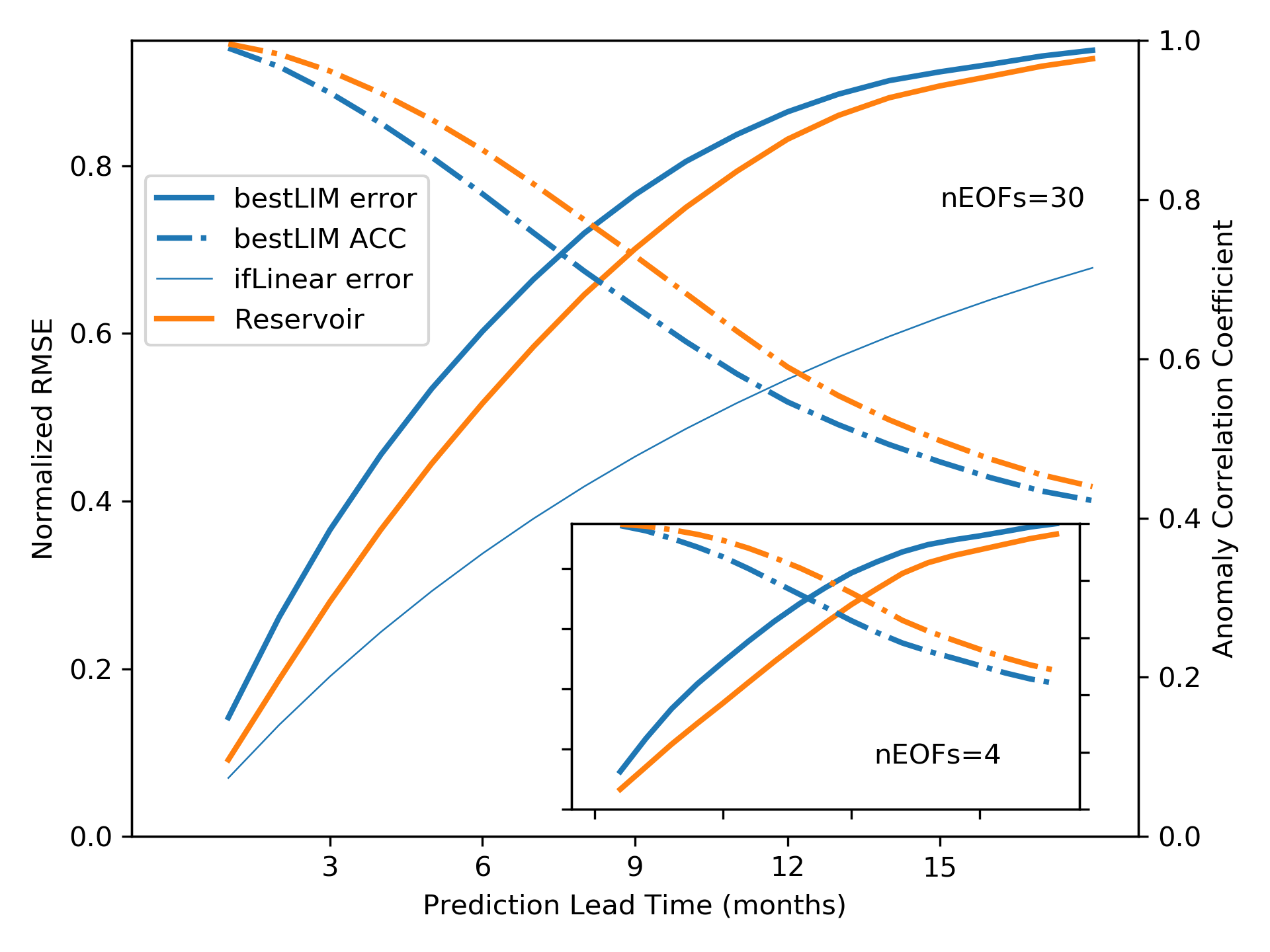}
  \caption{Spatially integrated skill measures are shown as a function
    of prediction lead time. Normalized RMSE is shown in thick lines
    using the left axis and anomaly correlation coefficient is shown
    in dot-dashed lines using the right axis. LIM results are shown in
    blue and RC results are shown in orange. The main plot shows
    results for predictions using 30 EOFs. The corresponding results
    for predictions using 4 EOFs is shown in the inset. All three axes in
    the inset have the same limits as in the main plot. The thinner
    blue line in the main plot is a theoretical estimate of the error
    in the LIM approach if the actual dynamics (that of SST in CESM2)
    were linear (notated as ifLinear in the legend; see (10)). The difference
    between the actual LIM error (thick blue line) and this
    theoretical estimate of error (ifLinear) is an indication of
    nonlinear dynamics in the data. The RC results show better skill,
    although not by much in this setting where training data is
    plentiful (480 years).}
  \label{erracclong}
\end{figure}

Predictions were limited to lead times $T_{pred}$ of 18 months since at that
time, the NRMSE had already reached values close to 0.9 and the ACC
had fallen to about 0.4 indicating the eventual approach to levels of skill
achieved by climatology. In order to make the comparisons between the
LIM and the RC approach statistically significant, predictions were
started every ($T_{skip}$) 18 months over the testing period and then statistics
were computed over such predictions (213 of them).

Figure~\ref{xyerrlong} shows the spatial distribution of NRMSE at a prediction
lead time of 12 months using the two methods, with the LIM result on
the left and the RC result on the right. First, the overall similarity
of the distribution of prediction error with the two methods indicates
overall similarity in the behavior of the two methods. It also
reiterates the greater predictability of the North Atlantic SST at
higher latitudes, in the region of the subpolar gyre as compared to
the mid and lower latitudes. Next, the smaller differences between the
two panels reveals that the prediction error with the RC methodology
is consistently lower than with the LIM approach.

Figure~\ref{xyacclong} shows the spatial distribution of the anomaly
correlation coefficient for the two methods again at the same
prediction lead time of 12 months as in Fig.~\ref{xyerrlong}. The
comparison of the two panels again reveals overall similarity, but
again with the RC predictions showing consistently higher correlation
to the reference ESM results.

To be able to better compare the skill of predictions using the two
methods, we next consider the spatially-averaged prediction error and
correlation as a function of prediction lead time. This is shown in
Fig.~\ref{erracclong}. In this figure, the LIM results (and diagnostics) are
shown in blue lines where as the RC related results are shown in
orange lines. The plots show NRMSE as a function of prediction lead time in heavy
solid lines and uses the left axis, and ACC likewise in dot-dashed
lines and uses the right axis. In addition, the thinner blue line
shows the theoretical estimate of error of the LIM if the actual
system were itself linear, and as given by (\ref{iflerr}). The
main plot is for predictions using 30 EOFs where as the inset plot is
for predictions using 4 EOFs. The inset plot uses the same limits for
the horizontal axis and for each of the two vertical axes as in the main plot.

The difference between the theoretical estimate of LIM error if the
actual system were linear and the actual LIM error indicates that
there is indeed nonlinearity in the actual system and provides a basis
for the expectation that nonlinear methods may be able to improve on
the linear approach of LIM. Indeed, this expectation is seen to be
realized by the simple RC model which essentially embodies a
``weighted sum and nonlinearity'' approach. While the improvement of
predictive skill of RC over that of LIM is small, it is
significant and consistent, both in terms of its variation with lead
time and at different number of EOFs. The significance of this
improvement is further reiterated on recalling that LIMs have been
speculated to capture the bulk of the predictable signal in
experiments that compare LIM predictions with predictions of full
dynamical models \citep{newman2017we}.

\subsection{Learning from Limited Data}
In the previous subsection we established the approximate similarity
of the LIM approach and the RC approach, but with the RC predictions
exhibiting better skill, when data is plentiful.
Physically, training over such long samples of data amounts to
generalizing the (18 month) response over a wide variety of
situations.
The dominance of linear behavior in that setting can be understood in
numerous ways including statistical mechanics (fluctuation-dissipation
theorem) and statistics. While it should be evident, we explicitly
note that this does not mean that the system itself is linear. To wit,
the commonplace recognition of the chaotic nature of climate (and
weather) on the short timescale is clear evidence of the nonlinearity
of the underlying dynamics. In this section, we are interested in this
regime, a regime where nonlinearity has a greater role to play. The
relevance of this regime to (initialized) interannual prediction
(cf. CMIP) cannot be overstated.

As such, we now focus on comparing the two methods when training data
is more limited: We limit the training period ($T_{tr}$) to about ten
years (11.5 to 13.3 years; see below). With 800 years of data, we
started the learning process every six years to have a large enough
number of samples to estimate skill statistics. In particular, as in
the previous section, the training period (now, of 10 years) was
followed by a validation period and a testing period of equal
lengths. The testing comprised of either three prediction starts over
a testing period of 3.3 years (to give a 60:20:20
training:validation:test split) or just one 18 month prediction. Even
as the second testing protocol places reduced emphasis on
generalization error in evaluating predictive skill, it was considered
for the reason that such single predictions are quite often of
significant interest. We note that a new LIM model and a new RC model
is learnt for each of the short (about 13 or 16.7 year) segments. As
before, the best lag for each LIM was determined over the
corresponding validation period, but this time only lags of 1, 2, 3,
and 4 months were considered. Thereafter, the LIM state matrix was
recomputed and the RC model trained (i.e., $\vv W_{out}$ was computed)
over the combined training and validation splits (as in Sec.~5.1).

\renewcommand{\prfx}{figs-submit2/200_10yr_18_18_r32la:1_le:1_sp:1_ri:0.1_sp:0.5_wi:0.01}
\begin{figure}
  \includegraphics[width=\textwidth]{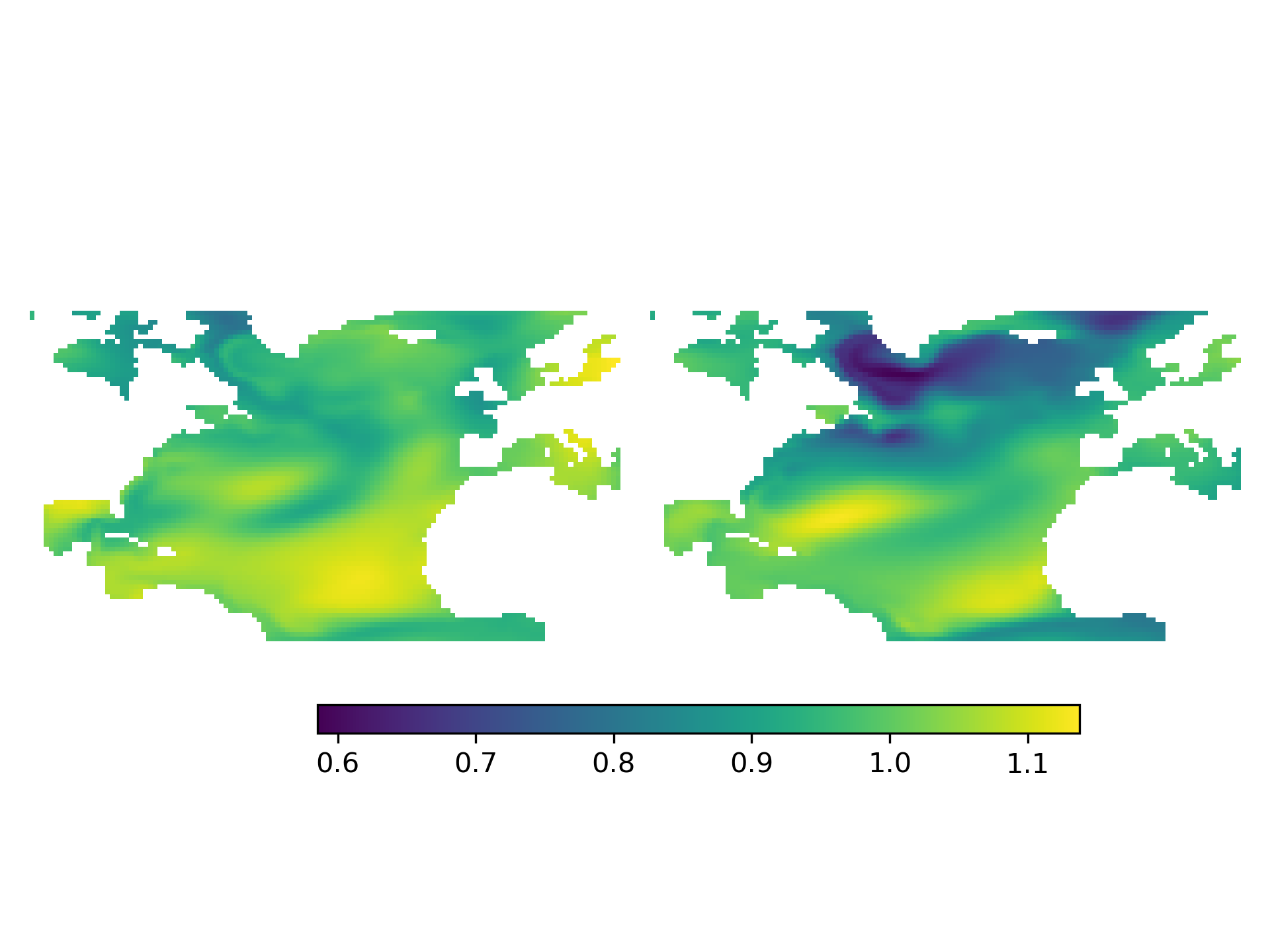}
  \caption{Comparison of NRMSE in the LIM predictions (left) and RC
    predictions (right), again at a lead time of one year (as in
    Fig.~\ref{xyerrlong}), but when training data is limited to about ten
    years. In this setting, the errors in the RC predictions are seen
    to be much smaller than in the LIM predictions.}
  \label{xyerrshort}
\end{figure}


\renewcommand{\prfx}{figs-submit2/200_10yr_18_39_r32la:1_le:1_sp:1_ri:0.1_sp:0.5_wi:0.01}  
\begin{figure}
  \includegraphics[width=\textwidth]{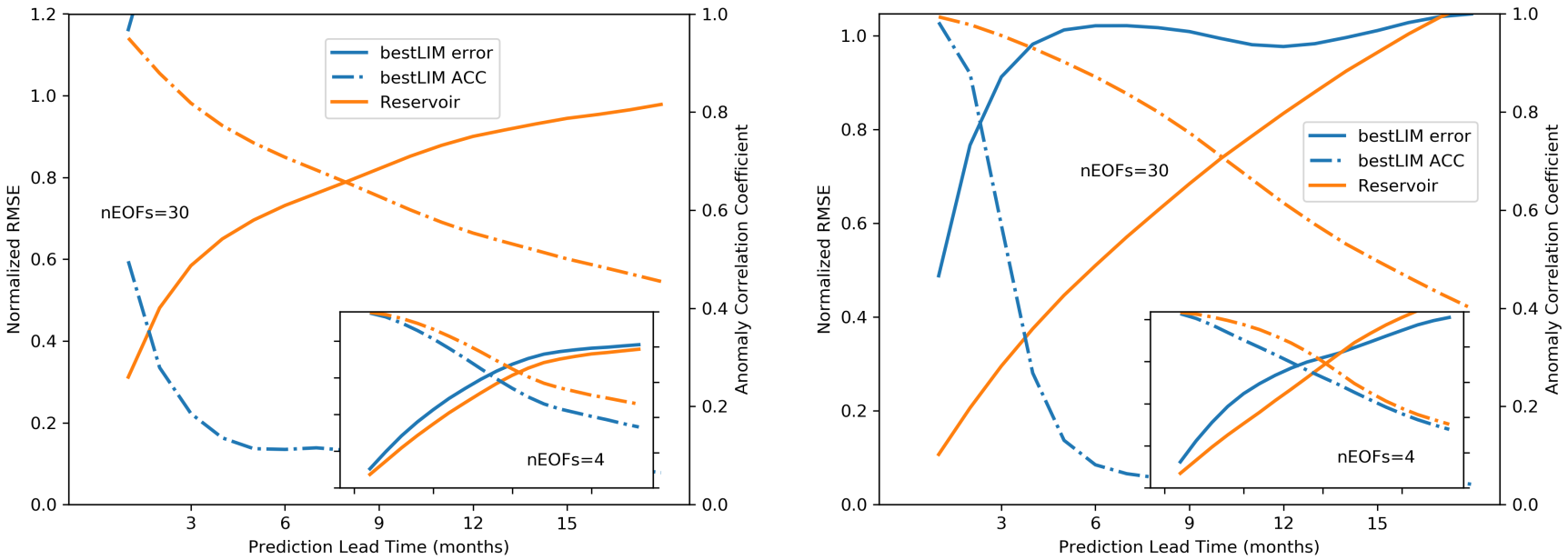}
  \renewcommand{\prfx}{figs-submit2/200_10yr_18_18_r32la:1_le:1_sp:1_ri:0.1_sp:0.5_wi:0.01}  
    \caption{Like in Fig.~\ref{erracclong}, this figure compares the
      spatially-integrated NRMSE (left axis) and ACC (right axis) for
      LIM predictions (blue) and RC predictions (orange) as a function
      of prediction lead time when training data is limited to about
      10 years. The main plots are for predictions that retain 30 EOFs
      whereas the inset plots are for predictions that retain only 4
      EOFs. The panel on the left corresponds to testing over (at
      least) 20\% of the data segment considered whereas the panel on
      the right shows results when testing is limited to predicting
      over a single 18 month period.  The RC predictions are seen to
      be more skilful than the LIM predictions when 30 EOFs are
      retained. When only 4 EOFs are retained, their skills are
      comparable.}
  \label{erraccshort}
\end{figure}

As alluded to earlier, the LIM approach can be problematic in this
setting in that the learnt system can be unstable. That is, one or
more of the eigenvalues of the estimated system matrix $\vv B$ can be
positive in the limited data setting, not because of shortcomings of
the estimation technique, but because the short runs of data are
indeed best explained in the linear framework by a system matrix that
has unstable eigenvalues. It seems that such an unstable system matrix
is prone to not generalizing well and leads to problems with
predictions. Rather than make ad-hoc adjustments to the procedure such
as limiting the number of EOFs considered or increasing the smoothing
of the data, other than noting this problem, we eliminate the LIM
prediction from computing skill measures when the prediction is
flagged to be unrealistic (e.g., if the predicted amplitude of any of
the EOF coefficients exceeded ten times the climatological standard
deviation over the prediction period). In so doing, the measures of
skill for the LIM approach we present will be (artificially)
inflated. For future reference, we also note that when greater than
50\% of the predictions are so flagged, skill measures are considered
unreliable and not used/presented (e.g., the leftmost points on the
blue curves in the main panel of Fig.~9 are omitted for this reason).

Using the same format as in Fig.~\ref{xyerrlong}, the error using LIM
and RC is shown in Fig.~\ref{xyerrshort} again at a prediction lead
time of 12 months (in the case where testing comprises of prediction
over a single 18 month period). While the spatial distribution of
errors is somewhat similar in the two methods in the sense that
errors at the lower latitudes are higher, etc., RC errors are
seen to be much lower than LIM errors. Similarly, the ACC for the RC
predictions are seen to be much larger than for the LIM predictions
(not shown). The variation of the spatially-averaged error and
correlation measures as a function of prediction lead time is shown in
Fig.~\ref{erraccshort} for the two testing protocols.
Again, blue lines are used for LIM results and
orange lines for RC results and the format is the same as in
Fig.~6. These results show that when learning from short stretches of data,
whereas the RC approach has much greater skill
than the LIM approach when the temporal evolution of the coefficients
of a large number of EOFs is being modeled (main plots in
Fig.~\ref{erraccshort}), their skill is comparable
when only very few (four) EOFs are retained (inset plots in
Fig.~\ref{erraccshort}). Practically speaking, this last point is somewhat of
academic interest alone. What we mean by that is: Since the skill of
the RC approach with 30 EOF coefficients is not much worse than
with only 4 EOF coefficients, even if the skill of the RC approach with 4
coefficients were to be worse or even much worse (than that of the LIM
approach), practically one would opt for the 30 EOF coefficient prediction
since it explains a much larger fraction of the variance at similar or
better skill.

\begin{figure}
  \includegraphics[width=\textwidth]{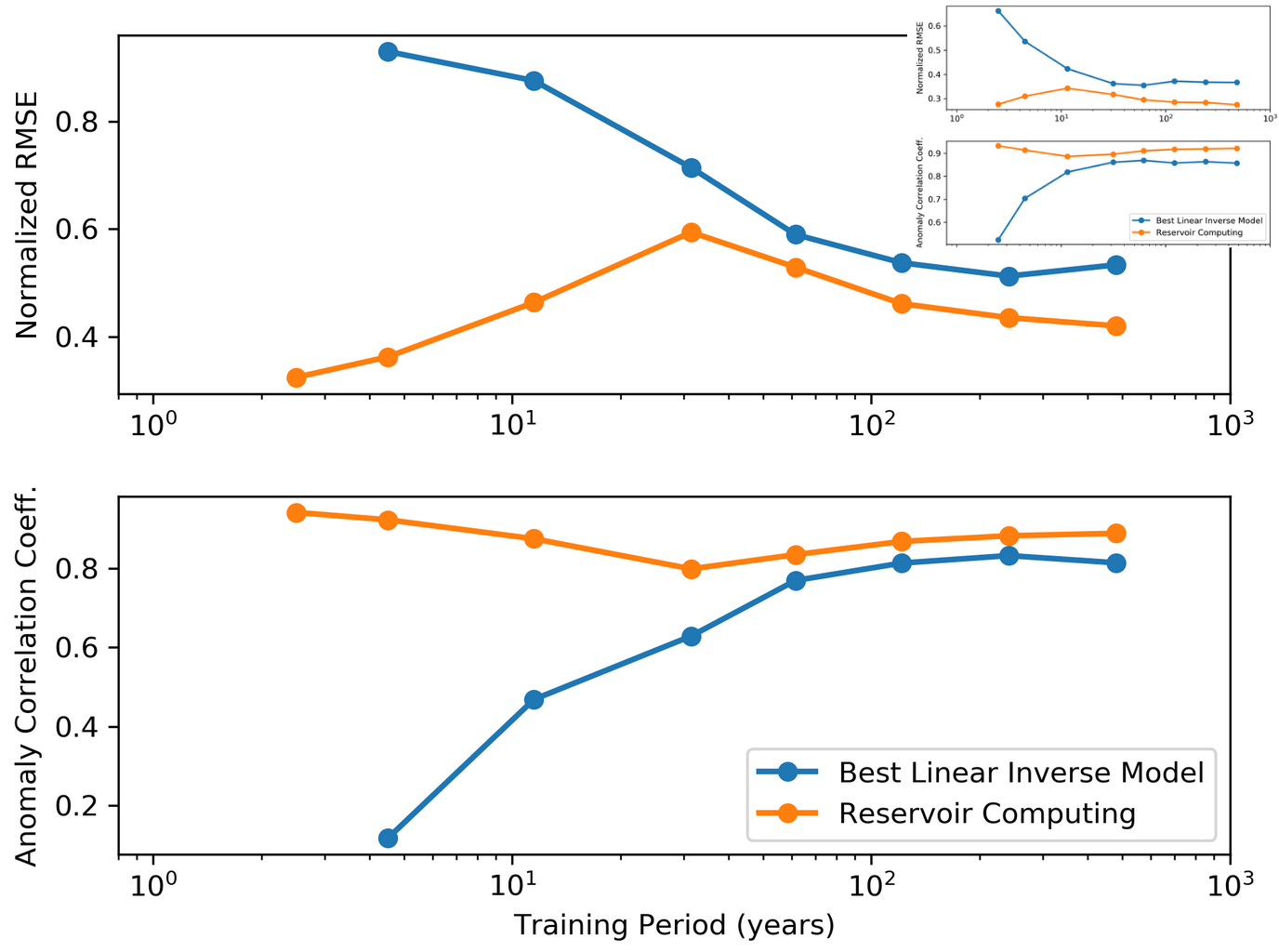}
  \caption{NRMSE and ACC of predictions of the year-1 averaged
    state. The main plots are for predictions retaining 30 EOF
    coefficients. The inset plots are for those retaining only
    4. While the LIM predictions display a monotonic behavior (more
    data, more accurate), the RC predictions show a distinct
    non-monotonic behavior with a regime where accuracy improves on
    limiting training data.}
  \label{err-vs-trn}
\end{figure}

On comparing Figs.~\ref{erraccshort} and \ref{erracclong}, it is
evident that when the RC approach outperforms the LIM approach it is
largely because of the degradation of skill of the LIM approach; the
skill of RC remains comparable when learning from long or short
stretches of data. If $N_y$ (practically the same as $N_u$) is the
number of EOFs retained, LIM in essence has to estimate
$N_y\left(N_y+1\right)/2$ parameters ($\vv C_{\tau_0}$ in (4)) over the training data whereas
the RC approach has to estimate a much larger number of parameters
that corresponds to the size of the $\vv W_{out}$ matrix
($\approx N_y\left(N_y+N_r\right)$, $N_r\gg N_y$). As such, one might
expect that limited data will have a greater deleterious effect on the
RC approach. Clearly, that is not the case.

At first glance, this behavior may be viewed in terms of RC being
successful in fitting a better local model since it allows for
nonlinearity. But, it is important to note that the better model
also generalizes since otherwise we wouldn't be seeing improvemnts in
skill over the test data. In addition, we suggest that it is this behavior---the
ability to perform well in situations where the number of parameters
that have to be learnt are far greater than the number of samples they
have to be learnt from---that the RC method shares in
common with other machine and deep learning techniques and which
constitutes an improvement over traditional statistical methods such as
LIMs.

In order to further verify this behavior, a number of other
experiments were conducted and Fig.~\ref{err-vs-trn} shows results
from one such experiment. In this figure, the NRMSE and ACC skill
measures for the prediction of the average state over the first year
is plotted as a function of the training period. The main plot is for
computations that retained 30 EOF coefficients whereas the inset plot is for
computations that retained only four. For the case of the shortest
period of training data, since greater than 50\% of the predictions
were flagged as unrealistic in the LIM approach, the skill measures
are not indicated for that case. Perusing the plots right to left, the
(more-or-less) monotonic degradation of accuracy of the LIM
predictions with decreasing training data is as would be expected from
a statistical perspective. The RC predictions, however, show a clear
departure from this expectation with a pronounced non-monotonic
behavior: skill of the RC predictions are worst at intermediate length
training data (around 30 years for 30 retained EOFs and around 10
years for 4 retained EOFs). While this aspect needs to be investigated
further and will be reported on in the future, we speculate that
this non-monotonic behavior is related to a transition from a regime
where the RC response is strongly non-linear when data is
limited to a regime where the RC response is more linear
when larger amounts of data are available for training.

\section{Discussion}
We considered a combination of settings, that included long and short
learning periods and when a small number of EOFs are retained (that
explained $\approx$ 50\% of variance) vs. when a larger number of EOFs
are retained (that explained $\approx$ 95\% of variance). In almost
all of these cases, the RC approach performs at least as well as the
LIM approach. For simplicity, these cases can be considered as falling
into two categories: one where training data is plentiful in
comparison to the number of parameters to be estimated and one where
data is more limited in that sense.  Whereas when learning data was
plentiful, the RC approach was only marginally better than the LIM
approach, the RC approach performed much better than the LIM approach in
settings of limited data. This difference in performance was mostly
related to the strong degradation of the performance of the LIM
approach. The predictive skill as a function of lead time
(prediction-horizon) of the RC approach in the limited data setting
was similar to that in the pentiful data setting. This needs to be
investigated further. Nevertheless, the reasonable performance of the
RC approach in the limited data setting is a significant and important
strength. Indeed, in a more general context, it is this related ability of
machine learning and deep learning models to generalize from limited
data that has led to their widespread popularity \citep[e.g.,
see][]{zhang2017understanding}. 

At this point, it is natural to wonder if the set of parameters used
were in some sense optimal and as to what the sensitivity of the
results is to changes in the parameters. As to the former, we recall
that a set of parameters found to work reasonably during an initial
exploration of the methodology in a particular setting was modified
minimally and used to carry out all the computations in the previous
section. For instance, if we consider the computation in which the RC
results show a larger NRMSE at lead times of about a year and longer
(inset of right panel in Fig.~8), with some searching, we were able to
find new parameter settings where NRMSE of the RC predictions were
everywhere lower. We avoided such a search mainly to keep the
presentation simple. As to the latter other than noting that such a
sensitivity analysis is beyond the scope of this article, considering
the different settings in which the same set of parameters has
performed reasonably, we are inclined to think that the qualitative
nature of the results are likely robust, and will not change, for
small but significant changes in the parameters. A limited range of
(opportunistic) testing that resulted, more often than not, to {\em not}
change the qualitative nature of the results presented is consistent
with our expectation.

\begin{figure}
  \includegraphics[width=\textwidth]{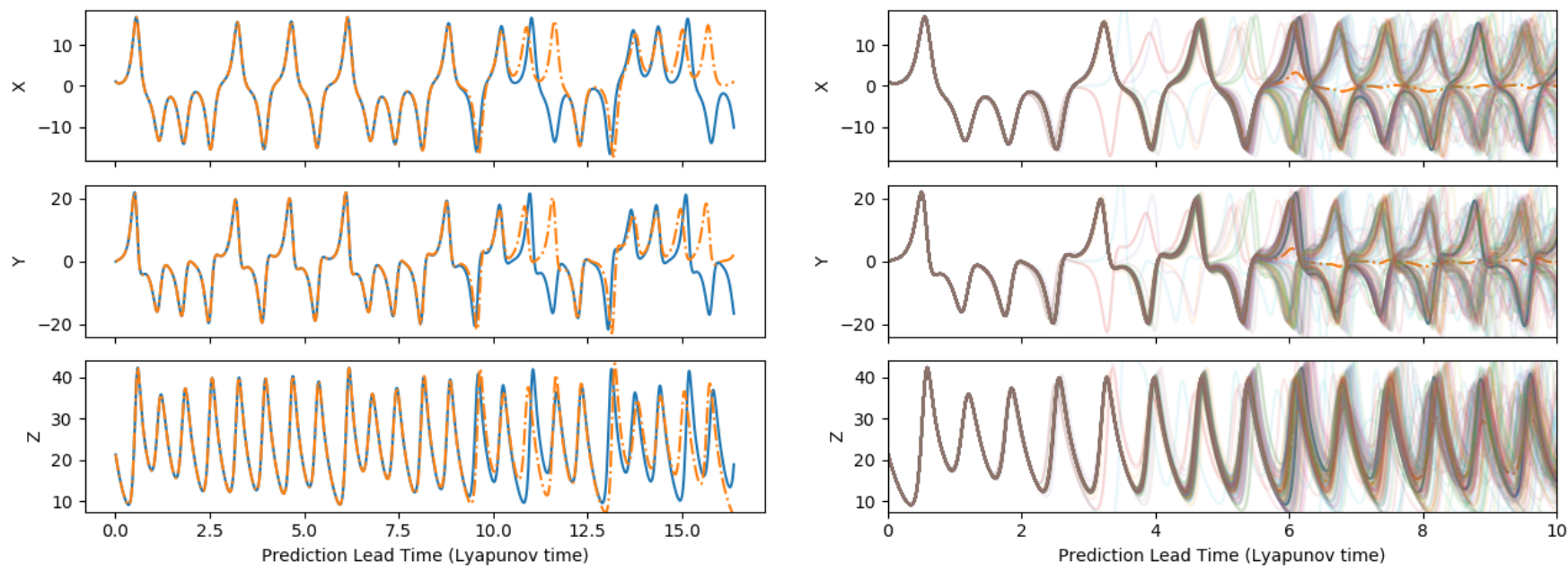}
  \caption{Reservoir computing based predictions of the Lorenz-63
    system. On the left is an individual prediction whereas the full
    ensemble (of 128 realizations) and the ensemble-mean are shown on
    the right for each of the three variables (rows). The reference
    trajectory is shown in blue (evident in the left column, but
    obscured by the ensemble mean or the ensemble-members in the right
    column). The prediction lead time on the x-axis is
    non-dimensionalized in terms of the Lyapunov time (an e-folding
    time for error growth).}
  \label{L63traj}
\end{figure}

\begin{figure}
  \includegraphics[width=\textwidth]{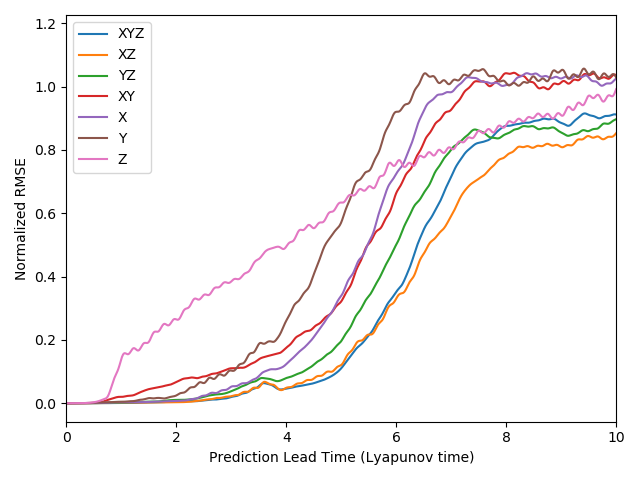}
  \caption{Normalized error as a function of prediction lead time for
    predictions using reservoir computing. In general (see body for
    details), the predictive skill diminishes when only progressively
    smaller parts of the system are observed and used in the RC
    learning and prediction procedure.}
  \label{L63error}
\end{figure}

\begin{figure}
  \includegraphics[width=\textwidth]{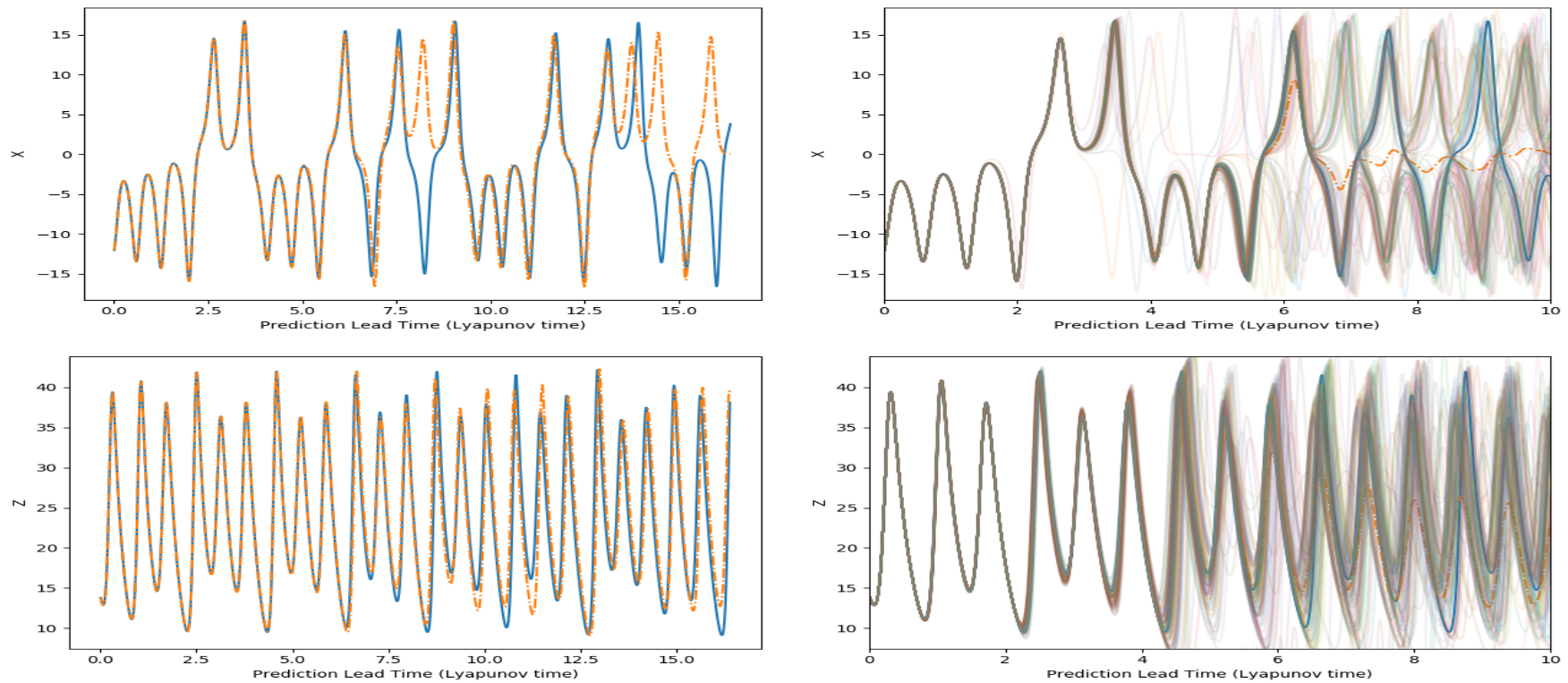}
  \caption{This figure follows the scheme in
    Fig.~\ref{L63traj}. Whereas Fig.~\ref{L63traj} was for RC
    predictions when the system was fully observed, this figure is for
    RC predictions when only one of the variables was used in the RC
    learning and prediction procedure. Top: RC learning and prediction
    using only the $x$ variable in (\ref{lorenz63}) (results are
    similar with $y$ alone). Bottom: with only $z$ variable.}
  \label{L63trajXZ}
\end{figure}

Machine learning methods span a wide range of complexity and it
remains an open question as to what methods are most suited for
climate predictability studies in settings such as the one we consider
and how far prediction skill itself can be improved. In this context,
we note that preliminary results in other on-going related work where
we consider other RNN architectures \citep{nadiga2019predicting,
  parkmachine, jiang2019interannual} suggest that comparable skill may
be obtained using techniques with more complexity such as
Convolutional Long Short Term Memory (convLSTM) architectures and that
the skill of less complex architectures such as Multi-Layer
Perceptrons (MLP) tends to be poor \citep[also
see][]{chattopadhyay2020data}.

On the other hand, and as mentioned earlier, it might be the case that
the reservoir computing paradigm that we have considered here is well
suited for the problem at hand in that they have proved exceptionally
good at tasks related to nonlinear system identification, prediction,
and classification in the context of chaotic dynamics. To wit,
Fig.~\ref{L63traj} compares the predicted evolution of an ensemble of
trajectories to a reference trajectory of the iconic Lorenz-63 system
\citep{lorenz1963deterministic}, hereafter referred to as L63:
\begin{eqnarray}
  \dot x &=& 28x - y - xz\cr
             \dot y &=& 10\left(y-x\right)\cr
                        \dot z &=& -\frac{8}{3}z + xy
                                   \label{lorenz63}
\end{eqnarray}  
Here, the three-dimensional state vector (or a subset of it) was
sampled every 0.01 time units for 100 time units after discarding an
initial integration period of 10 time units and learning of the
chaotic attractor and variability was achieved using the same
reservoir computing approach and parameters as described in the
previous section, but with a couple of modifications. The leakage parameter
$\alpha_{lk}$ in the reservoir update equation (\ref{resevolve}) is
now changed from the value of unity to 0.2 to approximately account
for the fact that the average decorrelation time in this setting is
about five times greater than that in the North Atlantic SST setting,
measured in terms of the respective time intervals between data
points.  Furthermore, since in the previous
setting we were trying to model one aspect in one region of a 
more complicated and more extensive multiscale system that was possibly sampled
infrequently, we might expect that regularization (related to ridge
regression) would be more important than in the  presently setting where we are dealing
with a toy three dimensional model that is well sampled. Indeed, on
trying a few smaller values for the $\alpha^2$ parameter in
(\ref{ridge}), we find that the predictions are better at a value of
10$^{-6}$ for the $\alpha^2$ parameter.
In other details, the data was split in a 60:40 train:test
fashion, predictions were continued for 18 time units and predictions
over the test section were started every 0.2 time units to allow for
110 start times. Performance measures were averaged over the 110
starts and the ensemble average of the predictions were performed over
128 realizations.

In both the panels of this figure, the x-axis corresponds to the
prediction lead time non-dimensionalized in terms of the Lyapunov time
($\approx$ 0.11 time units), a characteristic time related to error
doubling. The panel on the left shows a case where the RC is seen to
follow the reference trajectory over as long a period as ten times the
Lyapunov time.  The panel on the right shows the ensemble of RC
predictions for the same reference trajectory. While the ensemble mean
of the RC predictions (dot dashed orange) tracks the reference
trajectory till approximately 5.5 Lyapunov times, a number of the
ensemble members are seen to deviate earlier. It is also interesting
to note that at later times, after the ensemble mean has departed from
the reference trajectory, the structure of the attractor is still
seen to be well predicted. That is in the case of the Lorenz '63
system, the RC prediction system learns and predicts the structure of
the attractor well in addition to be being able to predict the exact
reference trajectory over significant periods of time (many Lyapunov
times).
Thus, like the results of \cite{pathak2018model}, Fig.~\ref{L63traj}
demonstrates the skill of RC, in being able to learn chaotic
variability.

We conducted a few more experiments to be able to relate the good
performance of the RC approach in the L63 context to the kinds of
performance that we presented in the previous section in the context
of predicting SST in the North Atlantic in the pre-industrial control
run of the CESM2 climate model. Figure.~\ref{L63error} documents the
results of these experiments in terms of the growth of error as a
function of prediction lead time when the system is fully observed
(blue lines) or only partially observed (other colors). Here, by
partially observed, we mean that the learning and prediction were
performed with only one or two of the three variable system (for a
total of six cases---the six colors other than blue). One set of RC
predictions when the system was fully observed (blue line) was
considered in Fig.~\ref{L63traj}. While we prefer the use of NRMSE as
a convenient diagnostic for various reasons, its usage in
Fig.~\ref{L63error} hides the fact that when different combinations of
variables are chosen for learning and prediction, the reference level
of climatological variance changes in each of the experiments.  It is
for this reason that the error in a partial-observation experiment
(e.g., the experiment labeled XZ) can show up as being less than in
the fully observed system.  Otherwise, a conclusion that may be drawn
from these experiments is that the skill in the RC predictions
decreases as fewer variables of the system are observed. That is to
say, the faster loss of skill in the SST context may likewise be due
to the fact that regional SST is only one small component of the, more
complicated, full model climate system that is being modeled in isolation.

Plots similar to Fig.~\ref{L63traj}---the fully observed case---are
shown in Fig.~\ref{L63trajXZ} for (one of the 110 sample predictions
each) for the two cases where only the X variable is observed (top
row) and only the Z variable is observed (bottom row). The case when
only the Y variable is observed is similar to when only the X variable
is observed and so not shown. Again, beyond times when the reference
trajectory is correctly predicted, the RC predictions are seen to
capture the climatological behavior reasonably. This is in spite of
the fact that the dimension of the observed and modeled system is too
low to even allow for chaotic behavior (cf. Poincare-Bendixon theorem,
\cite[e.g., see][]{guckenheimer2013nonlinear}). We conclude this
section by noting that we are working on further improvements to the
RC methodology that will permit us to use it similarly (as in the L63
setting) as a climate emulator.

\section{Conclusion}
Ensemble simulations are an essential tool for understanding
predictability of climate. Thus, when dealing with observations of the
actual climate system or with comprehensive climate models in their
own right, reduced-order dynamical models play a central role by
enabling simulation of large ensembles when such ensemble simulations
are otherwise either not possible or are computationally expensive and
resource-intensive. Furthermore such reduced order dynamical models
facilitate the testing and comparison of 
dynamical mechanisms across models and in observational data.

While the LIM approach has proved valuable as such a reduced order
modeling strategy, we have demonstrated that the nonlinear reservoir
computing approach exhibits useful skill over a wider range of
conditions including extending into regimes of limited data (more
nonlinear) and being able to accomodate large numbers of EOF
coefficients (that explain a greater fraction of the variance). While
such improved predictive skill of the new approach needs to be
verified in other settings, we do not see a priori reasons why this
should be difficult or impossible unless of course the system itself
is largely linear.  Clearly, establishing predictive skill of a method
is the first step towards facilitating its use in understanding issues
related to predictability such as the dynamics of error growth and the
structure of optimal perturbations, and we expect to report on these
topics in the near future.

The difference in computational cost between the LIM approach and the
RC approach is not a significant consideration in the reduced order
modeling framework that we have considered: In particular, given the
linear setting and the additive nature of the stochastic forcing in
the LIM approach, explicit averaging over noise realizations is not
required. However, the nonlinear nature of the RC approach
necessitates explicit averaging over random realizations of the
weights involved (but the realizations are trivially
parallelizable). Apart from that, a crude estimation of the asymptotic
scaling of the RC approach suggests an O($N^2_{\tilde r} T_{tr}$)
scaling for the plentiful data regime and O($N^3_{\tilde r}$) in the
very short training data regime, making it's computational cost very
low in the reduced dimension setting we have considered.

In addition to the previously mentioned difficulty of optimizing the
design of the reservoir, the issue of interaction of randomnes of the
weight matrices with other hyper-parameters needs to be better
understood. For example, in certain hyper-parameter regimes, while
some realizations of the random weights may give reasonable
predictions, others could be unreasonable. That is to say, the
ensemble can be over-dispersive. It is possible that such behavior is
more common when the reservoir is operating in a regime that is close
to violating the ``echo state property.'' Even though the latter
property is controlled by parameters such as the spectral radius of
the connectivity matrix, it is difficult to guarantee the property in
an a priori fashion while also guaranteeing reasonable performance. A solution
to this problem resides in part in tuning the system to be away from
such regimes using hyper-parameter tuning. A less-attractive 
alternative (because it modifies the ensemble spread explicitly) is to
address this issue in a posterior fashion as was 
done for the LIM approach in the limited data setting.

Aside from such practical considerations, at a more fundamental level,
given its basis in nonlinear kernel methods, it is possible that the
RC approach, will give insights into the nature of predictability that
are not accessible to linear points of view, where an example of an
insight provided by the LIM approach is that of non-normal finite time
amplification of perturbations in an asymptotically stable system
\cite[e.g., see][]{penland1995optimal, farrell2001accurate}.  For
example, apart from improved predictive skill, it is possible that
future developments of the RC method will permit its use as a climate emulator.

\section*{Acknowledgement}
We acknowledge the World Climate Research Programme, which, through
its Working Group on Coupled Modelling, coordinated and promoted
CMIP6. We thank the CESM2 modeling group for producing and making
available their model output, the Earth System Grid Federation (ESGF)
for archiving the data and providing access, and the multiple funding
agencies who support CMIP6 and ESGF. CESM2 SST data is publicly
available from the CMIP archive at
https://esgf-node.llnl.gov/projects/cmip6 and its mirrors. The author
was supported, by the Regional and Global Model
Analysis (RGMA) component of the Earth and Environmental System
Modeling (EESM) program of the U.S. Department of Energy's Office of
Science under the HiLAT-RASM project, by LANL’s LDRD
program, project number 20190058DR, and by DOE's SciDAC project ``Non-Hydrostatic Dynamics with Multi-Moment Characteristic Discontinuous Galerkin Methods''. We would like to thank the two
anonymous referees for their constructive criticism and comments; the
article has significantly benefitted from them.

\bibliography{main}

\end{document}